\documentclass[aps,superscriptaddress,nofootinbib,showkeys]{revtex4-2}
\usepackage{amssymb}
\usepackage{amsmath}
\usepackage{graphicx}
\usepackage[normalem]{ulem}
\usepackage{url}
\usepackage{csquotes}
\usepackage{xcolor}

\makeindex
\providecommand{\keywords}[1]{%
  \large	
  \textbf{\textit{Keywards:}} #1
}

\begin{document}

\title{Probing the density dependence of nuclear symmetry energy through isospin transport in heavy-ion reactions}
\author{\large S. Mallik}
\email{swagato@vecc.gov.in}
\affiliation{\mbox{Physics Group, Variable Energy Cyclotron Centre, 1/AF Bidhan Nagar, Kolkata-\textsf{700064}, India}}
\affiliation{\mbox{Homi Bhabha National Institute, Training School Complex, Anushakti Nagar, Mumbai-\textsf{400094}, India}}
\author{\large F. Gulminelli}
\email{gulminelli@lpccaen.in2p3.fr}
\affiliation{\mbox{Université de Caen Normandie, ENSICAEN, CNRS/IN2P3, LPC Caen, F-14000 Caen, France}}
\affiliation{\mbox{Institut Universitaire de France (IUF), France}}
\author{\large C. Ciampi}
\email{caterina.ciampi@ganil.fr}
\affiliation{\mbox{Grand Accelerateur National d'Ions Lourds (GANIL), CEA/DRF - CNRS/IN2P3, Caen, F-14076, France}}
\author{\large D. Gruyer}
\email{gruyer@lpccaen.in2p3.fr}
\affiliation{\mbox{Université de Caen Normandie, ENSICAEN, CNRS/IN2P3, LPC Caen, F-14000 Caen, France}}

\begin{abstract}
The density dependence of the nuclear symmetry energy remains one of the key uncertainties in contemporary nuclear physics, with significant implications for the structure of exotic nuclei, the dynamics of heavy-ion collisions, and the properties of astrophysical objects such as neutron stars and core-collapse supernovae. However, extracting robust constraints requires observables that are minimally affected by final-state interactions and are reliably predicted by transport models. This review synthesizes recent theoretical and experimental advancements in constraining the symmetry energy by leveraging isospin diffusion in heavy-ion reactions within the Fermi energy domain. Recent results from the INDRA-FAZIA collaboration, including isospin transport ratio data, and Boltzmann-Uehling-Uhlenbeck (BUU) transport model calculations are highlighted. Confidence regions for the symmetry energy are extracted from isospin transport ratios and isospin diffusion currents by utilizing state-of-the-art nuclear functionals, including both ab initio and phenomenological approaches, with a particular focus on the density regions probed by these experiments. The resulting constraints will aid future Bayesian studies of the nuclear equation of state and contribute to a more unified understanding of dense matter in both terrestrial experiments and astrophysical environments.
\end{abstract}
\maketitle
\section{Introduction}
\begin{figure}[!b]
\begin{center}
\includegraphics[width=0.7\textwidth]{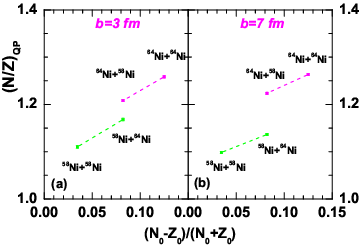}
\caption{Illustration of the evolution of the quasiprojectile neutron-to-proton ratio as a function of the initial isospin asymmetry for the $^{58,64}$Ni+$^{58,64}$Ni systems at two representative impact parameters, from a BUU@VECC-McGill transport simulation.}
\label{NbyZ_QP}
\end{center}
\vspace*{-.5cm}
\end{figure}

Understanding the properties of strongly interacting matter under extreme conditions remains a central challenge in modern nuclear physics. In this context, the nuclear equation of state (EoS) provides a quantitative description of the bulk properties of nuclear matter and determines its response to variations in baryon density and isospin asymmetry \cite{Bao-An-Li,Danielewicz}. The energy per nucleon of cold isospin asymmetric nuclear matter can be expressed as a function of baryon density $\rho=\rho_p+\rho_n$ and isospin asymmetry $\delta=(\rho_n-\rho_p)/(\rho_n+\rho_p)$. Within the parabolic approximation, it is written as
\begin{equation}
E(\rho,\delta)=E(\rho,\delta=0)+S(\rho)\delta^2+\mathcal{O}(\delta^4)
\label{EoS_Eq}
\end{equation}
where $E(\rho,\delta=0)$ denotes the isoscalar contribution and $S(\rho)$ is the symmetry energy. Introducing the dimensionless variable  $x=(\rho-\rho_0)/3\rho_0$, which measures the deviation of the baryon density from its saturation value, the density dependence of the isoscalar and isovector components around saturation density ($\rho_0$) may be expanded in powers of $x$. The normalization by $3\rho_0$ is adopted by convention, allowing the expansion coefficients to be directly related to the characteristic properties of nuclear matter at saturation density. The expansions read,
\begin{eqnarray}
E(\rho,\delta=0)=E_{sat}+\frac{1}{2!}K_{sat}x^2+\frac{1}{3!}Q_{sat}x^3+\frac{1}{4!}Z_{sat}x^4+...\nonumber\\
S(\rho)=E_{sym}+L_{sym}x+\frac{1}{2!}K_{sym}x^2+\frac{1}{3!}Q_{sym}x^3+\frac{1}{4!}Z_{sym}x^4+...
\label{EoS_term_Eq}
\end{eqnarray}
where, $E_{sat}$ is the isoscalar saturation energy, $K_{sat}$ characterizes the isoscalar incompressibility of symmetric matter, $Q_{sat}$ and $Z_{sat}$ represent higher-order density dependence (skewness and kurtosis, respectively). In the isovector sector, $E_{sym}$ is the symmetry energy at saturation, $L_{sym}$ corresponds to its slope, while $K_{sym}$, $Q_{sym}$ and $Z_{sym}$ describe the curvature and higher-order density dependence. The density dependence of the symmetry energy is of particular interest, as it plays a crucial role in a wide range of phenomena, such as the stability of neutron-rich nuclei \cite{Nature_review,Brown}, the dynamics of heavy-ion collisions \cite{Bao-An-Li,Danielewicz,Baran}, and the properties of dense stellar matter \cite{Lattimer_Physics_Report,Burgio,Sumiyoshi}. Unfortunately, the density-dependent symmetry energy $S(\rho)$ remains inadequately constrained, as theoretical predictions and experimental observations exhibit significant discrepancies across both the sub-saturation and supra-saturation regimes. In recent years, multimessenger observations, particularly gravitational-wave detections \cite{ligo_O1,ligo_O3,ligo_O3bis} from neutron star mergers have renewed interest in constraining the EoS, as it is essential for interpreting neutron star properties such as masses and radii \cite{Lattimer_NS_EoS}.\\
\indent
Despite significant progress, establishing a direct and unambiguous connection between experimental observables and the underlying symmetry energy remains challenging, primarily due to model dependencies and the complexity of reaction dynamics. Heavy-ion reactions, particularly those occurring near the Fermi energy, provide a unique opportunity to probe the symmetry energy and its density dependence \cite{Bao-an-li2,Dasgupta_book}. At these energies, nuclear matter is excited under conditions that partially overlap with those encountered in astrophysical environments, particularly in the sub-saturation density regime, enabling the study of nuclear matter at both sub-saturation and supra-saturation densities. During the reaction, the system undergoes compression followed by expansion, creating a dynamical environment in which the effects of the symmetry energy can be explored. Since the symmetry energy strongly influences the relative abundance of neutrons and protons, such reactions offer a powerful means to probe the EoS in the density regime relevant to neutron star interiors.\\
\indent
A variety of observables from heavy-ion reactions are sensitive to the nuclear equation of state, particularly to the symmetry energy, and provide important constraints over a wide density range. Among these, isospin transport \cite{Rami,Danielewicz2,Bao-an-li4,Tsang2004,Tsang_2009,yennello,Tsang2012,Zhang2020,Zhang2012} describes the redistribution of neutrons and protons during the collision and offers direct insight into the role of the symmetry energy in reaction dynamics. Isoscaling \cite{Tsang_isoscaling,Colonna,Mallik5,Mallik8,Ono,Mallik25}, which characterizes the isotopic distribution of fragments, reflects the influence of the symmetry energy on nuclear composition. The isobaric yield ratio \cite{Mallik8,Mallik25,Huang} offers additional constraints by comparing yields of isotopic pairs with different isospin. Furthermore, observables such as elliptic flow \cite{Russotto_flow,Trautmann_flow}, pion ratios ($\pi^-/\pi^+$) \cite{Bao-An-Li_pion,Zhigang_pion,Cozma_pion}, and neutron–proton differential flow \cite{Bao-An-Li_np_diff} provide complementary probes of the symmetry energy. In this work, we investigate the constraints on the nuclear equation of state obtained from isospin transport observables in intermediate-energy heavy-ion reactions, with particular emphasis on their sensitivity to the density dependence of the symmetry energy. A detailed understanding of isospin transport and its formalism is therefore essential for connecting experimental observables to the symmetry energy.\\
\begin{figure}[!h]
\begin{center}
\includegraphics[width=\textwidth]{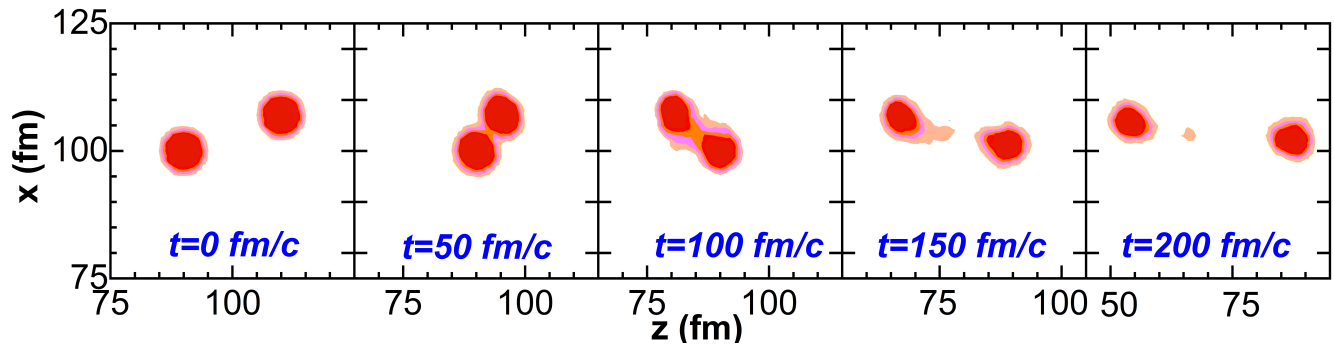}
\caption{Representative BUU@VECC-McGill transport simulation \cite{Mallik24} of the $^{58}$Ni+$^{64}$Ni reaction at 52A MeV and impact parameter $b=7$ fm. The snapshots illustrate the time evolution of the collision and the formation of a neck region that promotes isospin transport between the reaction partners.}
\label{Time_evolution}
\end{center}
\end{figure}
\indent
One of the important phenomena observed in heavy-ion collisions is the differential transfer of neutrons and protons in binary reactions, known as isospin diffusion. In the Fermi energy regime, where both mean-field effects and two-body collisions play significant roles, isospin diffusion manifests as a gradual redistribution of isospin between the colliding partners. During the overlap stage of the collision, a low-density neck region develops between the projectile and target nuclei, facilitating nucleon exchange and driving the system toward isospin equilibration. The extent of equilibration depends on the interaction time, the impact parameter, and the density dependence of the symmetry energy, making isospin diffusion a particularly sensitive probe of the isovector sector of the nuclear equation of state.\\
\indent
To illustrate the underlying transport mechanism, Fig. \ref{Time_evolution} presents a representative example of the collision dynamics. During the overlap stage, the formation of a neck region enables the exchange of nucleons between the reaction partners, thereby promoting the transport of isospin. As a consequence, the isospin content of the quasiprojectile and quasitarget evolves during the reaction toward a more equilibrated configuration. During the reaction, the projectile-like and target-like remnants survive as excited fragments commonly referred to as the quasiprojectile (QP) and quasitarget (QT), respectively. In the laboratory frame, the QP (QT) is identified as the fragment with the largest atomic number emitted in the forward (backward) direction relative to the beam axis. Since the QP is emitted in the forward direction, it can be detected in mass and charge by the FAZIA detector with high efficiency. Its isotopic composition, namely the neutron-to-proton ratio of the quasiprojectile, $(N/Z)_{\rm QP}$, provides a particularly useful probe of isospin transport and equilibration.\\
\indent
A direct manifestation of this process is shown in Fig. \ref{NbyZ_QP}, where the neutron-to-proton ratio of the quasiprojectile is displayed for the different $^{58,64}$Ni+$^{58,64}$Ni systems. The deviation of the mixed systems from the symmetric reference systems reflects the degree of isospin equilibration achieved during the collision. As expected, stronger equilibration is observed for smaller impact parameters owing to the increased overlap and interaction time between the reaction partners.

However, the neutron-to-proton ratio of the QP is strongly affected by secondary decay processes. In contrast, the isospin transport ratio (ITR) \citep{Rami} is constructed to minimize such effects through appropriate normalization, thereby providing a more robust probe of isospin diffusion. It is defined as
\begin{equation}
R=\frac{2x_{i}-x_{A_{l}+A_{l}}-x_{A_{h}+A_{h}}}{x_{A_{h}+A_{h}}-x_{A_{l}+A_{l}}}
\label{ITR_mathematical_form}
\end{equation}
where $x_i$ denotes an isospin-sensitive observable measured in the final state of the reaction.  Examples of $x_i$ include the neutron-to-proton ratio of the quasiprojectile \cite{Mallik24}, the neutron-to-proton ratio of free particles, or the isoscaling parameter \cite{Tsang2004}. The indices $A_l$ and $A_h$ refer to two colliding isotopes of the same element with different neutron contents, where $A_l$ is the less neutron-rich isotope and $A_h$ is the more neutron-rich one. The quantity $x_{i}$ corresponds to the observable measured in the mixed systems $(A_l+A_h)$ or $(A_h+A_l)$, while $x_{A_l+A_l}$ and $x_{A_h+A_h}$ are the corresponding values obtained for the two symmetric reference systems.\\
\indent
With this normalization, the symmetric reactions $(A_l+A_l)$ and $(A_h+A_h)$ correspond to $R=-1$ and $R=+1$, respectively. For the mixed systems, $R=0$ indicates complete isospin equilibration between the projectile and target, while values closer to $\pm1$ reflect weaker isospin transport and a stronger memory of the entrance-channel isospin asymmetry.\\
\indent
Pioneering studies by the MSU group at NSCL demonstrated that the sensitivity of $R$ to the symmetry energy arises from the dependence of isospin diffusion on the isospin chemical potential gradient \cite{Tsang2004}. This dependence enables constraints on the magnitude and slope of the symmetry energy around saturation density ($E_{\text{sym}}$ and $L_{\text{sym}}$). A stiffer symmetry energy generally leads to faster equilibration and smaller absolute values of $R$, whereas a softer symmetry energy preserves the initial isospin asymmetry \cite{Tsang2012}. However, precise constraints remain challenging due to uncertainties in comparing experimental data with transport model calculations. In particular, the isotopic composition of the QP residue was not directly measurable in earlier experiments, necessitating the use of surrogate observables such as isoscaling parameters and light cluster isobaric ratios \cite{Tsang2004,Tsang_2009}.\\
To address these challenges, the present article focuses on the following objectives:\\
\noindent\hspace*{0.5cm}(a) to examine the consistency of isospin transport ratios obtained from different isospin-sensitive observables;\\
\noindent\hspace*{0.5cm}(b) to quantify their sensitivity to the symmetry energy parameters $E_{sym}$, $L_{sym}$, and $K_{sym}$;\\
\noindent\hspace*{0.5cm}(c) to ensure that the selected observables are both experimentally accessible and reliably described by transport models \cite{Zhang2012}.\\
\noindent\hspace*{0.5cm}(d) to establish a robust and consistent determination of reaction centrality between experiment and theory;\\
\noindent\hspace*{0.5cm}(e) to determine consistently from the time evolution of the relevant currents the density region probed by the isospin transport ratio in order to avoid uncontrolled extrapolations;
and finally and most importantly;\\
\noindent\hspace*{0.5cm}(f) to reduce the uncertainties in constraining the density dependence of the symmetry energy  by implementing directly in the transport codes, and confronting to data, energy density functionals obtained by ab-initio calculations within the present uncertainties of chiral effective theory \cite{Drischler}.
\\
\indent
The experiments of $^{58,64}$Ni+$^{58,64}$Ni reactions at 32AMeV have been carried out at GANIL (Caen, France). Charged-particle multiplicities measured with the INDRA multidetector \cite{Pouthas1995, Pouthas1996} are used to reconstruct the impact parameter, while the isotopic composition ($N/Z$) of the QP remnant is directly measured using the high-resolution FAZIA telescope array \cite{Bougault2014,Valdre2019}. The correlation between these measurements provides a direct and largely model-independent probe of isospin diffusion. Theoretical calculations are performed within the BUU@VECC-McGill transport framework \cite{Mallik22,Mallik24,Mallik9}, based on the Boltzmann–Uehling–Uhlenbeck equation \cite{Huang_book,DasGupta_BUU_Physics_Report}, incorporating a metamodelling approach for the nuclear equation of state, that allows using complex realistic EoS functionals beyond the traditional simplified Zamick form \cite{Margueron2018a}.\\
\indent
The paper is organized as follows. Section \ref{Experiment} describes the experimental setup and data analysis. Section \ref{Model} presents the theoretical framework and model calculations. In Section \ref{Theory_Experiment}, the experimental results are compared with theoretical predictions to extract constraints on the symmetry energy. Finally, the conclusions are summarized in Section \ref{Conclusions}.

\section{Experimental Details}\label{Experiment}
The INDRA \cite{Pouthas1995, Pouthas1996} and FAZIA \cite{Bougault2014,Valdre2019} arrays are both optimized for the detection of charged fragments produced in heavy ion reactions at Fermi energies. They have been coupled in GANIL to exploit the large solid angle coverage of INDRA and the optimal identification performance of FAZIA, positioned at forward angles to cover the QP phase space \cite{Ciampi2022}. In the first experiment carried out with the coupled configuration, the four reactions $^{58,64}$Ni+$^{58,64}$Ni at 32AMeV have been measured to highlight the effect of isospin diffusion by comparing the QP remnants produced in the two asymmetric systems with those produced in the two symmetric ones. The isotopic composition of such heavy fragment is directly measured by FAZIA.

A key improvement of this work concerns the treatment of the reaction centrality, evaluated based on the charged-particle multiplicity measured by INDRA. The model-independent impact parameter reconstruction method of Ref.~\cite{Frankland2021} has been applied in order to extract the impact parameter $b$ distribution associated with each measured charged-particle multiplicity $M$: the method allows to retrieve both the correspondence between $M$ and $b$ and the associated intrinsic fluctuations. For the details of the specific implementation of such method for this work the reader is referred to Ref.~\cite{Ciampi2025}.

As isospin-sensitive observable we employ the $N/Z$ of the heaviest fragment produced in the forward hemisphere in the c.m. reference frame, directly identified by FAZIA and considered as QP remnant. For each event, we assign an impact parameter randomly drawn based on the $b$ distribution associated with the measured $M$, obtained model-independently as described above \cite{Frankland2021}: this procedure allows to take into account the fluctuations in the relationship between the two quantities, and to properly evaluate the model-independent average neutron to proton ratio $\langle N/Z\rangle$ of the QP remnant as a function of $b$. Finally, the isospin transport ratio is calculated with eq.~\eqref{ITR_mathematical_form} employing the QP remnant $\langle N/Z\rangle$ as isospin-sensitive observable: the resulting $R(\langle N/Z\rangle)$ as a function of the impact parameter $b$ is shown in Fig.~\ref{ITR_Theo_Expt}(a) with open rectangles \cite{Ciampi2025}. This model-independent experimental result can be easily compared with the predictions of transport models.

\section{Theoretical Approach: BUU@VECC-McGill Model}\label{Model}
In order to investigate the role of the nuclear equation of state in isospin transport, a consistent theoretical description of heavy-ion reaction dynamics is required. In this work, the BUU@VECC–McGill transport model is employed to simulate the dynamical evolution of the system at intermediate energies. The present study concentrates on observables in the positive rapidity domain, hence BUU@VECC–McGill transport model \cite{Mallik22,Mallik24,Mallik9} calculations are carried out in the projectile reference frame. The ground-state configurations of the projectile and target nuclei are constructed using a variational method \cite{Mallik9,Lee}, with the Myers density parametrization \cite{Myers} as the initial guess. The resulting ground-state density distributions are subsequently sampled using a Monte Carlo approach, where each nucleon is represented by $N_{test}=100$ test particles assigned appropriate spatial coordinates and momenta.\\
\indent
During the dynamical time evolution, the test particles with isospin $q=p,n$ propagate under the influence of a mean-field potential $U_q(\vec r,t)$ and may undergo binary nucleon–nucleon collisions during the time evolution. The probability of such collisions is determined by the isospin-dependent nucleon–nucleon cross section \cite{Cugnon}, provided that the final states are not forbidden by the Pauli blocking condition. The total mean-field potential is decomposed into bulk, surface, and Coulomb contributions as,\\
\begin{eqnarray}
U_{q}(\vec r,t)=U_{q}^{Bulk}(\vec r,t)+U_{q}^{Surf}(\vec r,t)+U_{q}^{Coul}(\vec r,t)
\label{Total_Mean_Field}
\end{eqnarray}
where $U_q^{Bulk}(\vec r,t)$ denotes the bulk contribution derived from the meta-functional introduced in Ref. \cite{Margueron2018a}. This functional is constructed through a polynomial expansion around the saturation density and incorporates deviations from the parabolic dependence on isospin asymmetry using density dependent isoscalar and isovector effective masses. The bulk potential is written as

\begin{eqnarray}
U_{q}^{Bulk}(\vec r,t)=(v^{is}_{0}+v^{iv}_{0}\delta^2)+\sum_{k=1}^{4}\frac{k+1}{k!}(v^{is}_{k}+v^{iv}_{k}\delta^2)x^{k}+\frac{1}{3}\sum_{k=1}^{4}\frac{1}{(k-1)!}(v^{is}_{k}+v^{iv}_{k}\delta^2)x^{k-1}\nonumber\\
+2\delta\tau_z(1-\delta\tau_z)\sum_{k=1}^{4}\frac{1}{k!}v^{iv}_{k}x^{k}+\exp\{-b(1+3x)\}\bigg{[}(a^{is}+a^{iv}\delta^2)
\bigg{\{}\frac{5}{3}x^4\nonumber\\
+(6-b)x^{5}-3bx^{6}\bigg{\}}+2\delta\tau_z(1-\delta\tau_z)a^{iv}x^{5}\bigg{]}
\label{Meta_model_potential}
\end{eqnarray}
where $x=\frac{\rho(\vec r,t)-\rho_0}{3\rho_0}$ and $\delta=\frac{\rho_n(\vec r,t)-\rho_p(\vec r,t)}{\rho(\vec r,t)}$. The coefficients $v_k^{is}$ and $v_k^{iv}$ ($k=1$–$4$) are connected to the isoscalar and isovecor empirical parameters described in Eq. \ref{EoS_term_Eq} by the relations given by,
\begin{eqnarray}
v^{is}_{0}=E_{sat}-t_0(1+\kappa_0)\;\;\;\;\;\;\;\;\;\;\;\;\;\;\;\;\;\;\;\;\;\;\;\;\;     v^{iv}_{0}=E_{sym}-\frac{5}{9}t_0[(1+(\kappa_0+3\kappa_{sym})]\nonumber\\
v^{is}_{1}=-t_0(2+5\kappa_0)\;\;\;\;\;\;\;\;\;\;\;\;\;\;\;\;\;\;\;\;\;\;\;\;\;\;\;\;\;
v^{iv}_{1}=L_{sym}-\frac{5}{9}t_0[(2+5(\kappa_0+3\kappa_{sym})]\nonumber\\
v^{is}_{2}=K_{sat}-2t_0(-1+5\kappa_0)\;\;\;\;\;\;\;\;\;\;\;\;v^{iv}_{2}=K_{sym}-\frac{10}{9}t_0[(-1+5(\kappa_0+3\kappa_{sym})]\nonumber\\
v^{is}_{3}=Q_{sat}-2t_0(4-5\kappa_0)\;\;\;\;\;\;\;\;\;\;\;\;\;\;\;\;\;v^{iv}_{3}=Q_{sym}-\frac{10}{9}t_0[(4-5(\kappa_0+3\kappa_{sym})]\nonumber\\
v^{is}_{4}=Z_{sat}-8t_0(-7+5\kappa_0)\;\;\;\;\;\;\;\;\;\;\;\;\;v^{iv}_{4}=Z_{sym}-\frac{40}{9}t_0[(-7+5(\kappa_0+3\kappa_{sym})]
\label{Isoscalar_parameters}
\end{eqnarray}
where $t_{0}=\frac{3\hbar^{2}}{10m}\Big{(}\frac{3\pi^{2}}{2}\Big{)}^{\frac{2}{3}}\rho_{0}^{\frac{2}{3}}$ is the kinetic energy per nucleon of symmetric nuclear matter, $m$ being the free nucleon mass. The parameters $\kappa_0$ and $\kappa_{sym}$ are related to the density dependence of the effective neutron mass $m^*_n$ and proton mass $m^*_p$ through the relations
\begin{eqnarray}
\frac{m}{m^*_n}=1+(\kappa_0+\kappa_{sym}\delta)\frac{\rho}{\rho_{0}}\nonumber\\
\frac{m}{m^*_p}=1+(\kappa_0-\kappa_{sym}\delta)\frac{\rho}{\rho_{0}}
\end{eqnarray}
The effective nucleon mass introduces an implicit momentum dependence in the interaction, which effectively accounts for the non-local character of the mean field at the intermediate beam energies of 30–60 MeV per nucleon explored in this paper. The values of the empirical parameters for the different EoSs used in this artile are listed in Table 1.\\
\begin{table}[!h]
\centering
\begin{tabular}{|c|c|c|c|c|c|c|}
\hline
Empirical & \multicolumn{6}{c|}{EoS} \\
\cline{2-7}
Parameter & Sly5 \cite{Sly5} & NL3 \cite{NL3_Paper} & SAMI \cite{SAMI_Paper} & SGII \cite{SGII_Paper} & \it{ab-intio 1} \cite{Drischler} & \it{ab-intio 7} \cite{Drischler}\\
\hline
$\rho_0$ (fm$^{-3}$) &0.1604& 0.1480 & 0.1587 & 0.1583 & 0.1890 & 0.1400\\
$E_{sat}$ (MeV) &-15.98& -16.24& -15.93 & -15.59 & -16.92 &  -13.23\\
$K_{sat}$ (MeV) & 230& 271 & 245 & 215 & 241 & 214 \\
$Q_{sat}$ (MeV) & -364& 198 & -339 & -381 & -125 & -139\\
$Z_{sat}$ (MeV) & 1592& 9302 & 1331 & 1742 & 1281 & 1306\\
$E_{sym}$ (MeV) & 32.03& 37.35 & 28.16 & 26.83 & 34.57 & 28.53\\
$L_{sym}$ (MeV) & 48.3& & 43.70 & 37.6 & 48.50 & 48.10 \\
$K_{sym}$ (MeV) & -112 & 101 & -120 & -146 & -224 & -172\\
$Q_{sym}$ (MeV) & 501& 182 & 372 & 330 & -311 & -164\\
$Z_{sym}$ (MeV) & -3087& -3961 & -2179 & -1891 & -1974 & -2316\\
$\kappa_{0}$ & 0.43 & 0.54 & 0.47 & 0.27 &0.70 & 0.49\\
$\kappa_{sym}$ & 0.18 & -0.40& -0.02 & -0.22 & -0.55 & -0.42\\
\hline
\end{tabular}
\caption{Saturation density $\rho_0$, isoscalar saturation energy ($E_{sat}$), isoscalar incompressibility ($K_{sat}$), isoscalar skewness ($Q_{sat}$), isoscalar kurtosis ($Z_{sat}$), symmetry energy at saturation ($E_{sym}$), symmetry energy slope ($L_{sym}$), isovector incompressibility ($K_{sym}$), isoscalar skewness ($Q_{sym}$), isoscalar kurtosis ($Z_{sym}$) and isoscalar effective mass parameter ($\kappa_{0}$) and isovector ($\kappa_{sym}$) effective mass parameter. See
text for more details.}
\label{tab:tab1}
\end{table}
\indent
The infinite nuclear matter functional is supplemented by a finite-range term in order to account for surface effects that arise in finite nuclei and heavy-ion collisions. Physically, this contribution penalizes rapid spatial variations of the density and therefore governs the properties of the nuclear surface and the interface between regions of different densities. From a theoretical perspective, such a term emerges naturally from the semi-classical $\hbar$ expansion of the non-local momentum-dependent interaction \cite{Lenk,Mallik10} and is represented by the surface energy density $U_q^{surf}(\vec r,t)=A\,\rho\,\nabla^2\rho$.
The corresponding coupling constant is fixed to $A=c/2\rho_0^{5/3}$, with $c=-6.5$ MeV. The Coulomb contribution is written as $U_q^{coul}(\vec r,t)=\frac{1}{2}(1-\tau_q)U_c(\vec r,t)$, where $\tau_q=-1$ for protons and $\tau_q=1$ for neutrons. The propagation of test particles in the mean field is performed using the lattice Hamiltonian method, which ensures an accurate conservation of both energy and momentum \cite{Lenk}. A comparison of observables obtained with the BUU@VECC–McGill model and other transport approaches based on BUU and quantum molecular dynamics can be found in Refs. \cite{Hermann,Zhang_code_comparison,Ono_code_comparison,Colonna_code_comparison}.\\
\indent
To account for cluster formation in heavy-ion reactions, transport calculations must be performed on an event-by-event basis, including fluctuations in the mean field \cite{Bauer}. In the present work, a computationally efficient prescription is adopted \cite{Mallik10,Mallik14,Mallik18}. Within this approach, nucleon–nucleon collisions are evaluated at each time step using isospin-dependent cross sections, but only among the $A_p+A_t$ test particles belonging to the same event. When a collision between two test particles $i$ and $j$ is allowed, the $(N_{test}-1)$ test particles closest to particle $i$ in coordinate space are identified, and the same momentum transfer $\Delta \vec{p}$ experienced by $i$ is assigned to all of them. Similarly, the $(N_{test}-1)$ test particles closest to particle $j$ receive the momentum change $-\Delta \vec{p}$. This procedure is continued throughout the time evolution of the event and repeated independently for each event. Details of the fluctuation-added BUU calculation used in the present work can be found in Section II of Ref.~\cite{Mallik10}. The nucleon–nucleon cross sections employed in the simulations correspond to free-space values parameterized from experimental data. Finally, fragments are identified using a spatial coalescence criterion: two test particles are considered to belong to the same cluster if their mutual distance does not exceed 2 fm \cite{Mallik14}. The metamodelling approach Eqs.(\ref{Meta_model_potential},\ref{Isoscalar_parameters}) with a density expansion up to fourth order allows a precise reproduction of realistic equations of state functionals obtained with phenomenological or ab-initio nuclear interactions.\\
\indent
The qualitative features illustrated in Figs. \ref{Time_evolution} and \ref{NbyZ_QP} are now investigated quantitatively within the BUU@VECC-McGill framework. For the initial theoretical analysis, we consider the $^{64,58}$Ni+$^{64,58}$Ni reactions from mid-central to very peripheral collisions and employ, as a benchmark interaction, the EoS parameter set fitted to the SLy5 functional \cite{Sly5}. At a later stage, the calculations are extended to equations of state derived from chiral effective field theory \cite{Drischler}, conventional Skyrme interactions, and relativistic mean-field models. The simulations are performed with 100 test particles per nucleon, and 500 events are generated for each reaction system. The $^{58,64}$Ni+$^{58,64}$Ni systems are selected for isospin diffusion studies because they involve stable isotopes with significantly different neutron contents, while being sufficiently heavy for bulk nuclear matter properties to play a dominant role in the reaction dynamics. The quasiprojectile (QP) is identified through its momentum distribution, which peaks near zero in the QP frame.\\
\indent
The QP freeze-out time is estimated from the time evolution of the momentum-space isotropy in the QP frame \cite{Mallik22,Mallik18}. Although complete equilibration is not achieved, the momentum distribution becomes approximately isotropic at around 100 fm/$c$, largely independent of the impact parameter and the projectile–target combination. In conventional analyses, the dynamical evolution is often followed by a statistical evaporation stage \cite{Camaiani,Mallik11} to extract final observables. However, this procedure may introduce additional uncertainties related to the choice of coupling time, excitation energy, and possible inconsistencies between the transport model and the statistical description. To avoid these ambiguities, the present work does not include an afterburner stage; instead, the time dependence of the isospin transport ratio is analyzed directly, and care is taken to consider times long enough for the observable of interest to reach its asymptotic value.
\subsection{Time Dependence of ITR from different observables}
\begin{figure}[!h]
\begin{center}
\includegraphics[width=0.8\textwidth]{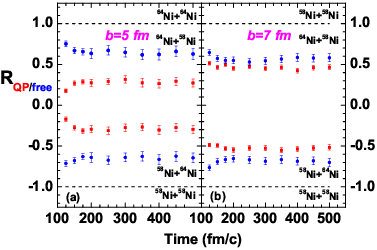}
\caption{Variation of the ITR with time for the $^{58,64}$Ni+$^{58,64}$Ni reactions at 52 MeV/nucleon, for impact parameters $b=5$ fm (left panel) and $b=7$ fm (right panel). Red squares denote ITR determined from the quasiprojectile, while circles correspond to forward-emitted free nucleons.}
\label{ITR_time_dependence}
\end{center}
\end{figure}

Within a transport model based on one-body dynamics, two different correlated observables can be constructed, with $x$ defined in Eq. \ref{ITR_mathematical_form}): (i) $x=N/Z$ of the quasiprojectile (QP), which defines $R_{QP}$, and (ii) $x=N/Z$ of the free nucleons emitted forward in the QP reference frame, which defines $R_{free}$. The former is expected to be a relatively robust mean-field observable, since it is only indirectly affected by the neck dynamics and the associated cluster production, which remain model dependent in transport calculations. For the latter, only nucleons with $p_z>0$ MeV/c in the QP frame are selected to reduce the contamination of free nucleons originated from the participant region and pre-equilibrium emission. Fig. \ref{ITR_time_dependence} displays the time evolution of $R_{QP}$ and $R_{free}$ for $^{64,58}$Ni+$^{64,58}$Ni reactions with projectile beam energy 52 MeV/nucleon at impact parameter $b$=5 and 7 $fm$. Both $R_{QP}$ and $R_{free}$ are found to remain nearly constant for $t\ge$150 $fm/c$, confirming a weak sensitivity to secondary decay effects. Indeed, independent GEMINI++ \cite{Charity,Mancusi} calculations  show that, while the residue $N/Z$ is sensitive to the residual
excitation energy that might still be stored in the QP when the calculations are stopped at a (short) finite time, this dependence is strongly weakened
when the ITR is constructed \cite{Mallik24}. Accordingly, the transport calculation is stopped at $t\ge$300 $fm/c$, and the values of $R$ averaged over $t=$150-300 $fm/c$, with a step of 50 $fm/c$, are used in the subsequent analysis.
\subsection{Entrance Channel effect on ITR}
\begin{figure}[!h]
\begin{center}
\includegraphics[width=0.8\textwidth]{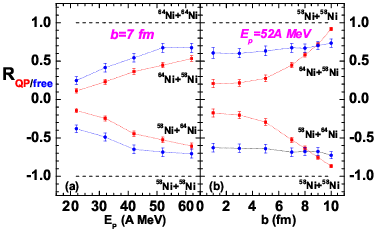}
\caption{Dependence of the isospin transport ratio on projectile energy (left panel) and impact parameter (right panel). The ratios extracted from the quasiprojectile neutron-to-proton ratio and from forward-emitted free nucleons are shown by red squares and blue circles, respectively. The projectile-energy dependence is evaluated at $b=7$ fm, whereas the impact-parameter dependence is presented for 52 MeV/nucleon. Dashed lines are included to guide the eye.}
\label{ITR_Entrance_Channel_dependence}
\end{center}
\end{figure}

The influence of the entrance-channel conditions on isospin diffusion, extracted from both the quasiprojectile asymmetry observable $R_{QP}$ and the free nucleon observable $R_{free}$, is illustrated in Fig. \ref{ITR_Entrance_Channel_dependence}. The left panel presents the dependence on projectile energy at a fixed impact parameter of $b=$7 $fm$, whereas the right panel shows the variation with impact parameter at a projectile energy of 52A MeV. The calculations indicate that the values of $R_{QP}$ and $R_{free}$ are not identical in general. As the collisions become more central, the participant zone increases in size, which promotes stronger isospin equilibration in reactions involving isospin-asymmetric systems. With increasing projectile energy, the magnitude of the isospin transport ratio increases for both observables. This trend can be attributed to the reduced interaction time, the growing contribution of nucleon–nucleon collisions, and the diminished role of the mean field. These observations suggest that experiments performed at lower beam energies might be
more suitable for extracting accurate information on the nuclear equation of state.
\subsection{Sensitivity of ITR on isovector parameters}
\begin{figure}[!h]
\begin{center}
\includegraphics[width=\textwidth,keepaspectratio=true]{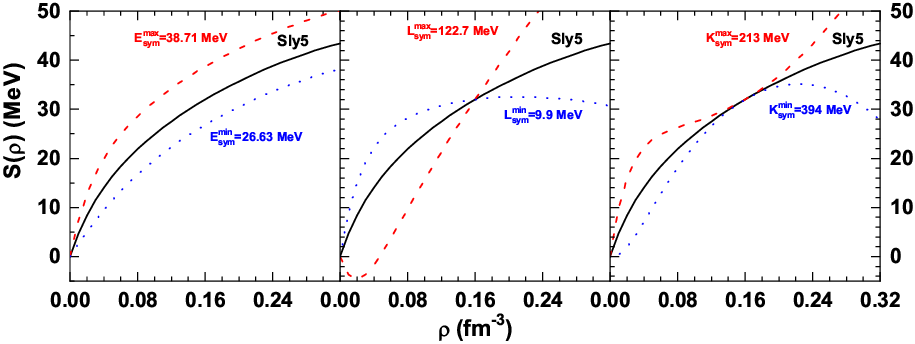}
\caption{Density dependence of the symmetry energy corresponding to schematic EoS models in which the isovector parameters $E_{sym}$, $L_{sym}$, and  $K_{sym}$ are varied independently (left, middle, and right panels respectively), while the remaining parameters are fixed to their SLy5 values.}
\label{Symmetry_energy_Esym_Lsym_Ksym}
\end{center}
\end{figure}

\begin{figure}[!t]
\begin{center}
\includegraphics[width=0.7\textwidth]{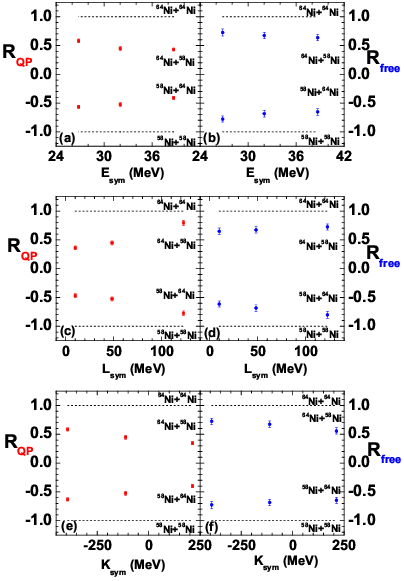}
\caption{ Dependence of the isospin transport ratio on the symmetry energy parameters $E_{sym}$, $L_{sym}$ and $K_{sym}$, for observables constructed from the neutron-to-proton ratio of the quasiprojectile (red squares connected by dashed lines) and from forward-emitted free nucleons (blue circles connected by dotted lines).}
\label{ITR_Esym_Lsym_Ksym_dependece}
\end{center}
\end{figure}

To investigate the sensitivity of the isospin transport ratio to the density dependence of the nuclear symmetry energy, at first the three lowest-order isovector parameters, namely the symmetry energy at saturation $E_{sym}$, its slope $L_{sym}$, and curvature $K_{sym}$, are varied independently. In this procedure, two of the parameters are fixed at their reference values corresponding to the SLy5 equation of state, while the remaining parameter is systematically varied over a range compatible with current empirical and theoretical constraints. The ranges considered for $E_{sym}$, $L_{sym}$, and $K_{sym}$ correspond to the minimum and maximum values of the empirical parameter intervals reported in the meta-modelling analysis of Ref.~\cite{Margueron2018a}. The resulting density dependences of the symmetry energy are illustrated in Fig.~\ref{Symmetry_energy_Esym_Lsym_Ksym}, where these limiting values are explicitly indicated. The corresponding effects of these variations on the isospin transport ratio for the $^{64,58}$Ni+$^{64,58}$Ni reactions at an impact parameter $b=$7 $fm$ and projectile energy of 52A MeV are presented in Fig. \ref{ITR_Esym_Lsym_Ksym_dependece}.\\
\indent
At this beam energy, the reaction dynamics predominantly probe densities around or below the saturation density. In this density region, the symmetry energy becomes smaller for lower values of $E_{sym}$ and $K_{sym}$ , as well as for higher values of $L_{sym}$. Since isospin transport is driven by the symmetry energy, a smaller symmetry energy leads to weaker isospin equilibration between the reaction partners. This reduced equilibration results in larger absolute values of the isospin transport ratio. Consequently, the magnitude of the transport ratio increases with increasing  $L_{sym}$, whereas it decreases with increasing values of $E_{sym}$ and $K_{sym}$. As a result, even if all low-order empirical parameters affect the isospin transport ratio, the sensitivity is found to be strongest for the slope parameter $L_{sym}$.
It is also noteworthy that equations of state such as those denoted $L_{sym}^{min}$ and   $K_{sym}^{max}$ produce similar isospin transport behavior at sub-saturation densities, despite exhibiting large differences at suprasaturation densities. This emphasizes that extrapolations from sub-saturation constraints to the high-density regime should be treated with caution. Because of these ambiguities, when extracting EoS constraints from nuclear observables, it is not enough to find the EoS models compatible with the data. Instead, one has to simultaneously infer the highest likelihood EoS model (or equivalently, the highest likelihood set of EoS parameters) and the density intervale probed by the experiment. This issue will be handled in Section \ref{sec:probe}. Both observables, $R_{QP}$ and $R_{free}$, exhibit sensitivity to variations in the symmetry energy parameters; however, the dependence is significantly more pronounced when the quasiprojectile-based observable is considered, indicating that it provides a more sensitive probe of the symmetry energy in this energy regime. It is worth noting that Ref.~\cite{Zhang2015} also identified the isospin transport ratio as a sensitive probe of the symmetry-energy slope parameter $L_{sym}$, while reporting additional sensitivities to the isoscalar nucleon effective mass and a moderate dependence on the isovector effective mass splitting. In the present work, the effective-mass parameters are kept fixed, and the analysis is focused on the sensitivity to the symmetry-energy empirical parameters within the BUU@VECC--McGill framework.\\
\indent
The above theoretical analysis indicates that the neutron-to-proton ratio of the QP is more sensitive to the isovector parameters than that of free nucleons. Also, it is important to recall that composite particles emitted by the QP together with free nucleons constitute the experimental samples, while one-body transport models interpret all emitted particles as free nucleons. In principle, one can compare $R_{free}$ to the analogous ratio obtained summing up the proton and neutron content of the different evaporated light charged particles and neutrons, but this would require the extra assumption that this ratio is coalescence invariant. An alternative path consists in building up an $R_{free}$ variable also theoretically, using a trasport model that treats clustering, such as a molecular dynamics code. However, dynamical cluster production is still subject to huge uncertainties that might affect the extraction of EoS information. On the other hand, the isotopic resolution of the FAZIA apparatus allows a direct comparison between model and data through $R_{QP}$, with a strong reduction of systematics arising from the treatment of clustering and/or the coupling to afterburners. These arguments all go in the direction of  focussing on $R_{QP}$. Concerning the incident energy, at relatively low projectile energies, (i) the mean-field effect is significant compared to nucleon–nucleon scattering; (ii) supra-saturation densities are not extensively probed, which reduces the mixing of information between sub-saturation and supra-saturation density regions; and (iii) the effect of secondary decay is 
smaller.
Based on these different arguments, in the next section we will focus on quantifying the density dependence of the symmetry energy from the ITR of the QP in $^{64,58}$Ni on  $^{64,58}$Ni reactions at 32 MeV/nucleon only.\\
\begin{figure}[!b]
\centering
\includegraphics[height=7cm,keepaspectratio=true]{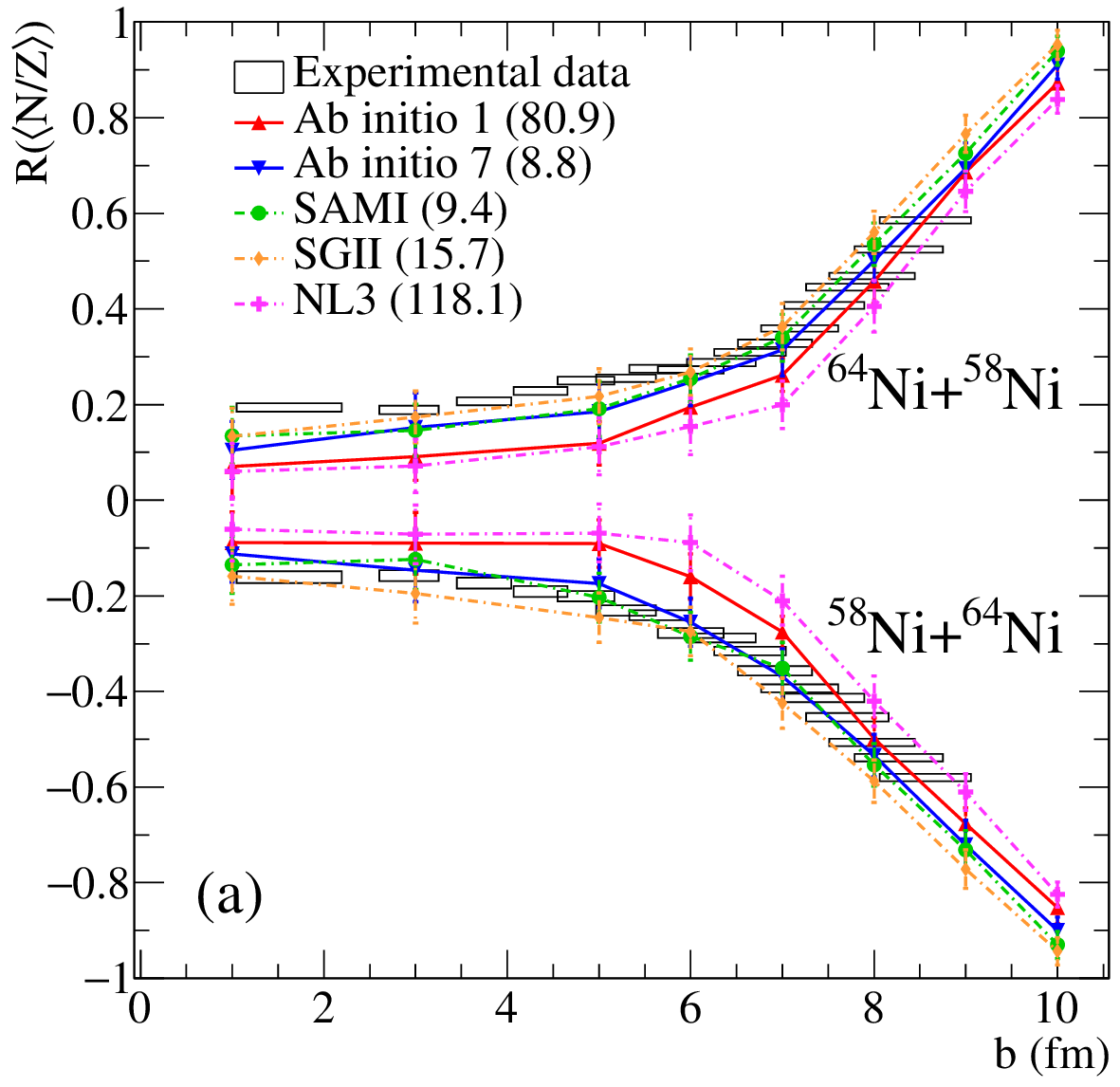}\qquad
\includegraphics[height=7cm,keepaspectratio=true]{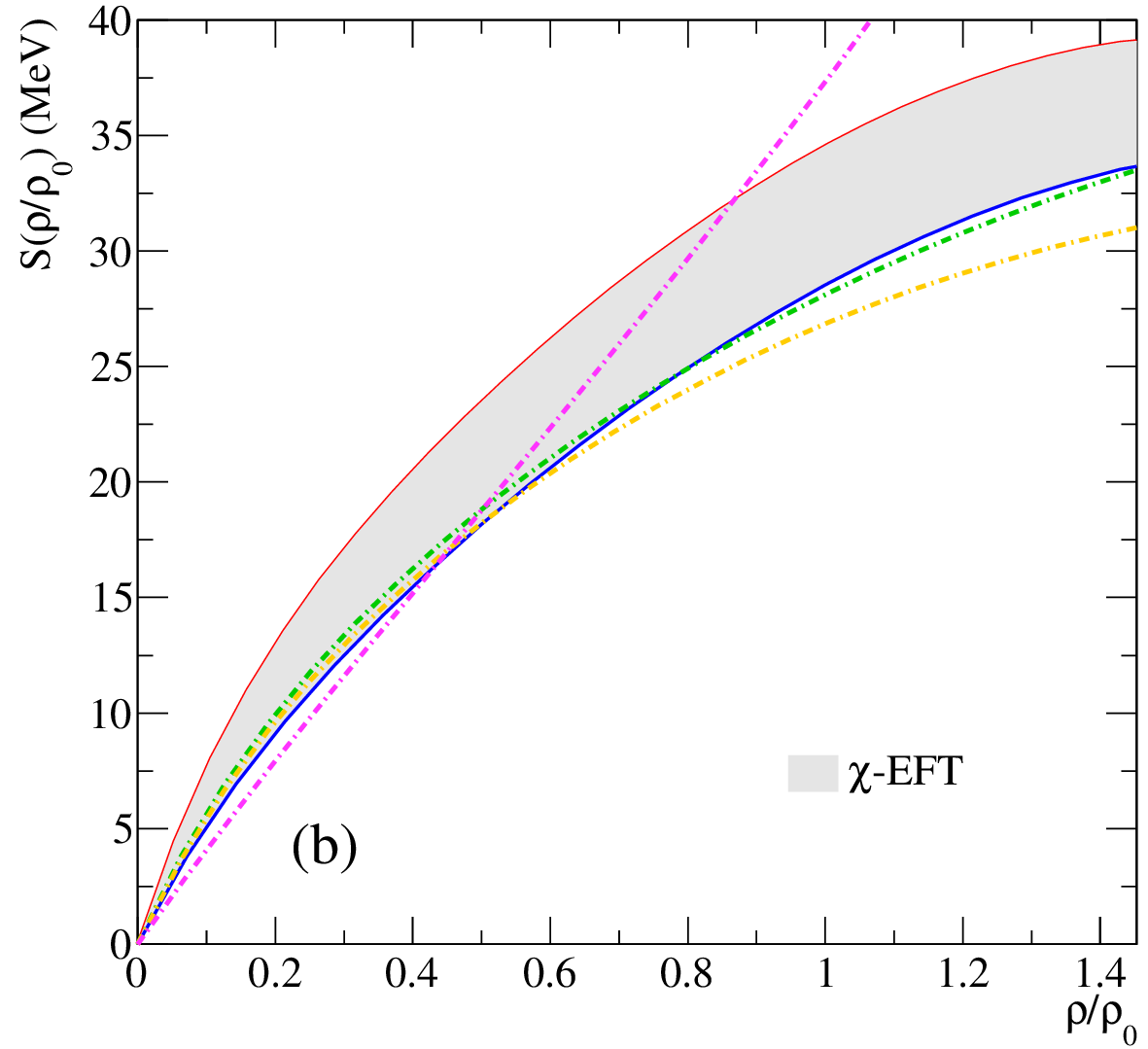}
\caption{Variation of the isospin transport ratio $R$ with impact parameter, determined from the quasiprojectile average $\langle N/Z \rangle$, for the $^{64}$Ni+$^{58}$Ni (upper increasing curves) and $^{58}$Ni+$^{64}$Ni (lower decreasing curves) reactions at a beam energy of 32 MeV per nucleon (left panel). The experimental measurements \cite{Ciampi2022} are displayed as open rectangles, with their widths indicating the corresponding uncertainties. These results are compared with predictions from the BUU@VECC–McGill transport model using different nuclear equation-of-state parametrizations, including \textit{ab initio}~1, \textit{ab initio}~7, SAMI, SGII, and NL3.  Calculations have been carried out for selected values of impact parameter, indicated with markers, and the curves are shown to guide the eye. The $\chi^2$ values associated with each parametrization are listed in the legend.\\
\indent
Right panel presents the density dependence of the symmetry energy for the same set of models, adopting a consistent color and line-style convention. Solid lines correspond to \textit{ab initio} calculations, while dash-dotted lines represent phenomenological approaches. The shaded gray region indicates the uncertainty band derived from chiral effective field theory constraints.
}
\label{ITR_Theo_Expt}
\end{figure}

\section{Symmetry Energy Constraints from Combined Experimental and Theoretical Analysis}\label{Theory_Experiment}
The sensitivity analysis of the lowest-order isovector parameters on isospin diffusion, as presented in Section 3.3, is not sufficient to provide quantitative predictions. This limitation originates from the fact that, within realistic equations of state (EoS), the empirical nuclear matter parameters are not independent but are subject to intrinsic correlations imposed by the underlying energy density functionals. As a consequence, varying individual parameters independently does not correspond to physically consistent EoS models. In order to evaluate the discriminative power of isospin diffusion with respect to realistic equations of state and to enable a meaningful comparison with experimental data, the isospin transport ratio (ITR) is calculated using several representative EoS parameterizations within the BUU@VECC–McGill model, including two extreme $\chi$-EFT interactions (models 1 and 7 in Ref. \cite{Drischler}, hereafter referred to as \textit{ab-initio-1} and \textit{ab-initio-7}), upon which the $\chi$-EFT confidence band for pure neutron matter is typically extracted for EoS inference studies. We also consider the Skyrme interactions SAMI \cite{SAMI_Paper} and SGII \cite{SGII_Paper}, and the relativistic mean-field interaction NL3 \cite{NL3_Paper}. These parameterizations are based on widely used nuclear energy density functionals that span the current uncertainties in the density dependence of the symmetry energy presented in the right panel of Fig. \ref{ITR_Theo_Expt}. In this way, the sensitivity of isospin transport observables to different realistic behaviors of the symmetry energy is investigated and their compatibility with experimental constraints is evaluated over the entire impact parameter range for  $^{64,58}$Ni on  $^{64,58}$Ni reactions at 32 MeV/nucleon. A fully consistent Bayesian analysis in principle requires a simultaneous variation of all equation-of-state parameters, including both isovector and isoscalar quantities, within their empirical uncertainty ranges. However, it is well known in the literature that the transport ratio observable is not  sensitive to the isoscalar part of the energy functional \cite{Coupland_2011}. The insensitivity of $R$ to a variation of isoscalar parameters was moreover explicitly checked for the present collision system in our transport model. The ITR predicted by the BUU@VECC–McGill calculations for the interaction models described above are shown in the left panel of Fig. \ref{ITR_Theo_Expt}, using the same color and line-style scheme as in the right panel. The error bars associated with the model calculations represent statistical uncertainties originating from the finite number of simulated events. The different interaction models considered in this study lie within the relatively narrow range of symmetry energy behaviors allowed by current experimental and theoretical constraints, and therefore lead to similar values of the ITR with a comparable dependence on the impact parameter. In particular, the very good agreement observed between the \textit{ab initio}~7 and SAMI functional indicates that the ITR is indeed sensitive to, and therefore capable of probing, the density dependence of the nuclear symmetry energy. Although the differences among the various theoretical predictions are relatively small, a systematic trend can still be identified. In particular, the models characterized by a comparatively large symmetry energy around saturation density, such as \textit{ab initio}~1 and NL3, can be ruled out, in agreement with the findings reported in previous studies \cite{Bao-An-Li_ITR}. The best overall agreement between the calculated ITR and the experimental data is obtained for the SAMI, SGII, and \textit{ab initio}~7 interactions. These models are characterized by similar values of the symmetry energy in the density region relevant for isospin transport, which explains their comparable performance in reproducing the experimental observations.\\
\indent
The level of agreement between the transport model calculations performed with different equations of state and the experimental data is quantified through the $\chi^2$ measure defined as,
\begin{equation}
\chi^2_{EoS}=\frac{\sum_{b=b_{\text{min}}}^{b_{\text{max}}}[R_{ex}(b)-R_{th,EoS}(b)]^2}{2[\sigma_{ex}^2(b)+\sigma_{th,EoS}^2(b)]}
\label{Chi_sq}
\end{equation}
In this expression, the summation runs over the selected range of impact parameters, $R_{ex}(b)$ and $R_{th,EoS}(b)$  denotes the experimental values and corresponding theoretical predictions with particular EoS for the ITR respectively. The total variance is defined as $\sigma_{ex}^2(b)+\sigma_{th,EoS}^2(b)$, thereby accounting for both experimental and theoretical uncertainties. This statistical estimator provides a quantitative measure of the deviation between theoretical predictions and experimental observations over the considered impact parameter range. In the present analysis, the summation is performed over impact parameters ranges from $b_{\text{min}}=3$~fm to $b_{\text{max}}=9$~fm, corresponding to the region where the experimental data are most sensitive and where the isospin transport observable provides the strongest constraints on the equation of state.\\
\indent
The $\chi^2_m$ values obtained for the different parametrizations considered in this work can be used to estimate approximate confidence regions in the $S$--$\rho/\rho_0$ plane. For each density bin $\rho_i$, the posterior probability distribution of the symmetry energy $S_i \equiv S(\rho_i)$ is constructed as
\begin{align}
p_i(S_i|ITR)= \mathcal{N} \sum_{EoS} p_i(EoS|ITR)\; \delta \left (S_i-S^{EoS}(\rho_i) \right ),
\end{align}
where the sum runs over the models included in the analysis, $S^{EoS}(\rho_i)$ denotes the symmetry energy predicted by particular EoS at density $\rho_i$, and $\mathcal{N}$ is a normalization constant. The EoS posterior probability $p_i(EoS|ITR)$ is obtained from Bayes’ theorem assuming a Gaussian likelihood,
\begin{align}
p_i(EoS|ITR)=p^{prior}_i\exp{(-\chi^2_m/2)}.
\end{align}
A flat prior is assumed for the symmetry energy in each density bin, such that $p^{prior}_i = w(\rho_i/\rho_0)$, where the weight function reflects the density region effectively probed by the experiment.\\
\begin{figure}[!h]
    \centering
    \includegraphics[height=5cm]{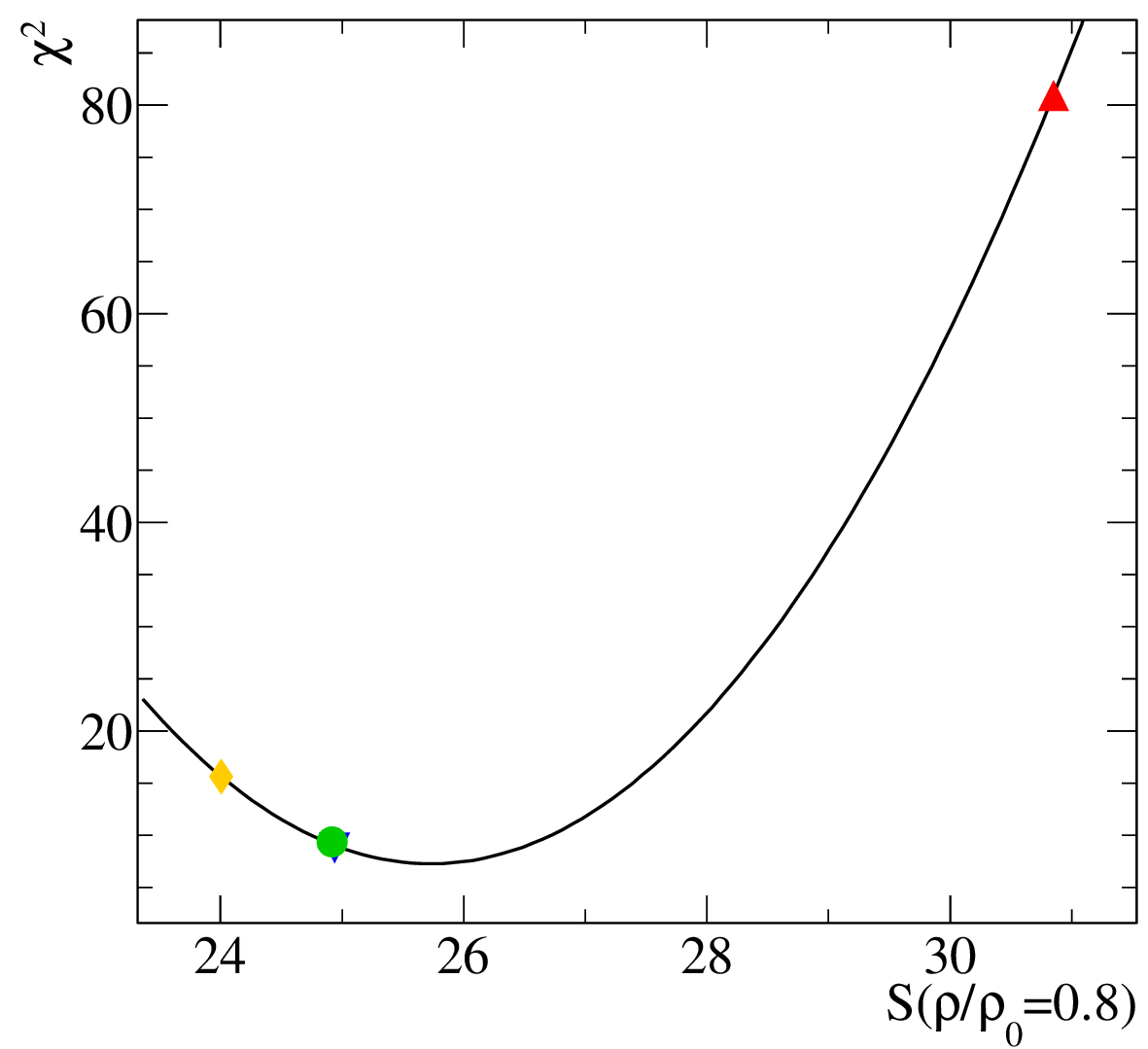}\quad
    \includegraphics[height=5cm]{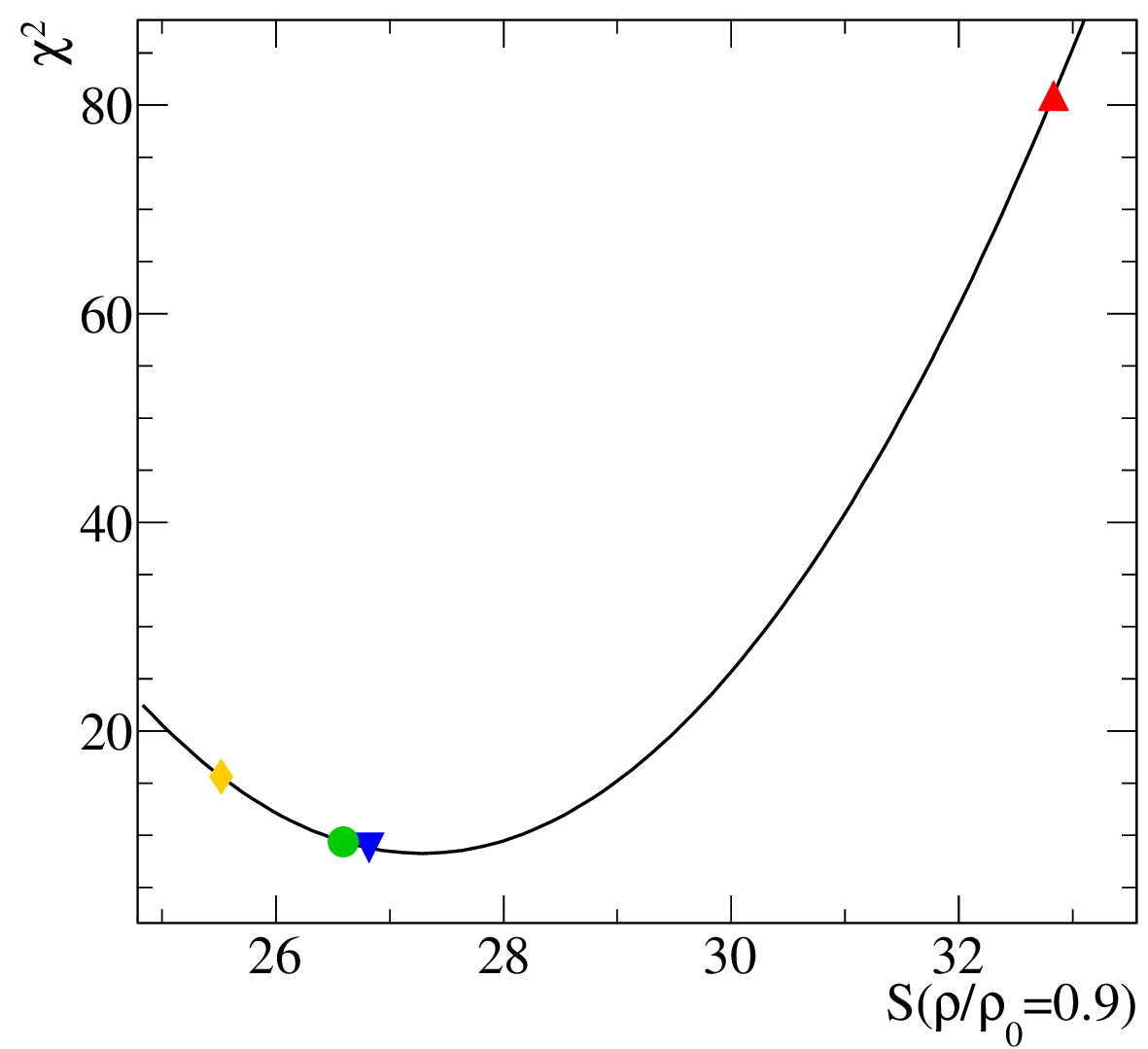}\quad
    \includegraphics[height=5cm]{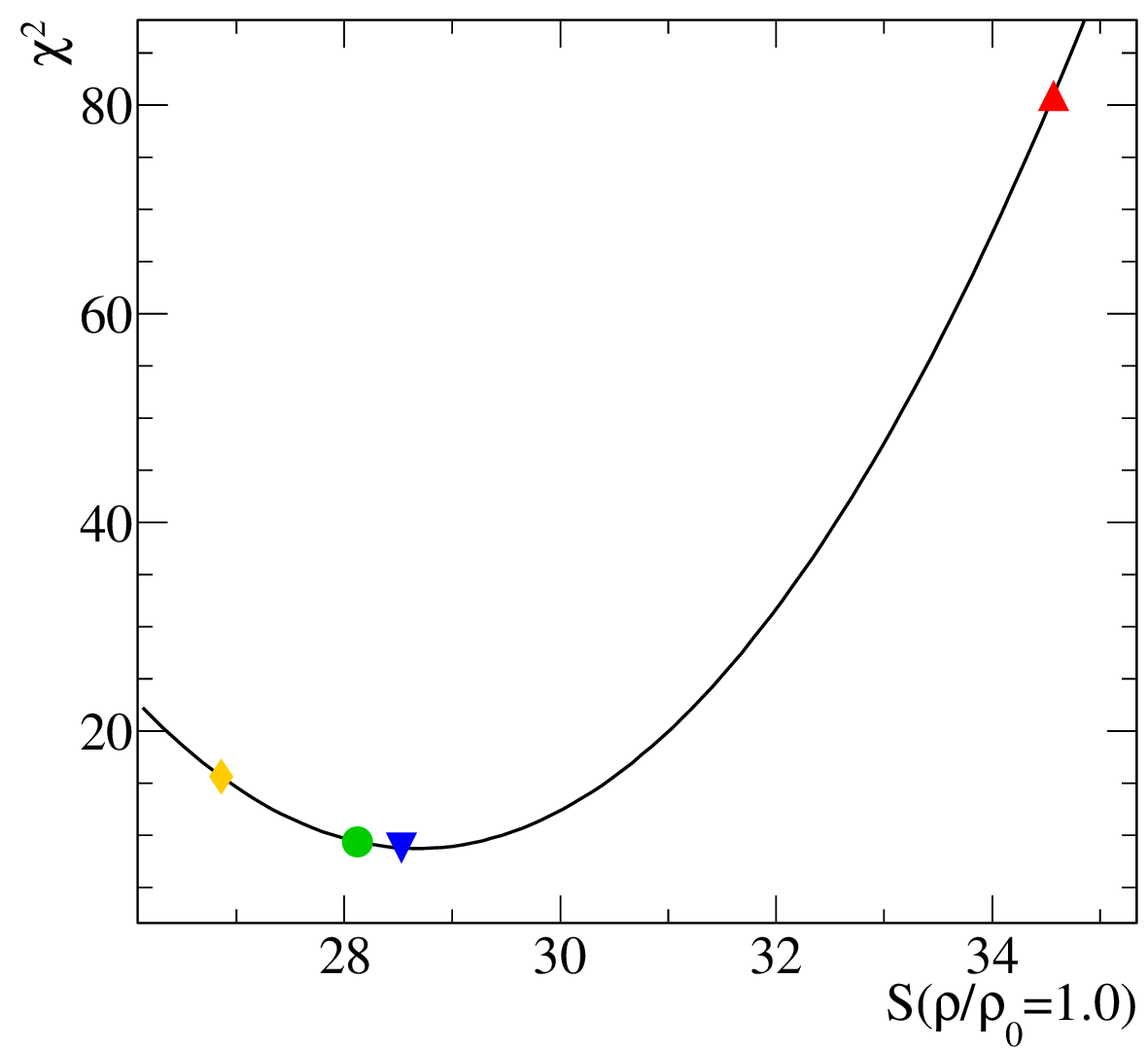}
    \caption{Examples of the $\chi^2$ values as a function of $S(\rho/\rho_0)$ for three selected densities, illustrating the quality of the quadratic approximation used in the analysis. The marker colors correspond to the different nuclear equation-of-state parametrizations following the legend of Fig.~\ref{ITR_Theo_Expt}. The black line shows the quadratic fit of $\chi^2(S(\rho/\rho_0))$, as an example of the procedure.}
    \label{fig:chi_fit}
\end{figure}
\indent
In principle, the $\chi^2$ values should be evaluated for a large set of models covering the full prior range of the symmetry energy, which would require a large number of transport calculations, which would not be computationally
accessible without a massive use of emulators. However, in the density range $0.4 \lesssim \rho/\rho_0 \lesssim 1.1$, the $\chi^2_m$ values show an approximately parabolic dependence when plotted as a function of $S(\rho)$. Therefore, as a first step toward a full Bayesian exploration of the parameter space, the likelihood for intermediate symmetry energy values is obtained through a quadratic interpolation between the selected models. The quality of this approximation can be appreciated from Fig.\ref{fig:chi_fit}, which shows the $\chi^2$ value of the different models from Eq.(\ref{Chi_sq}), plotted as a function of the value of the model symmetry energy at different densities. We can see that the quadratic dependence is well verified for a large density interval, with the exception of
the NL3 parametrization, which shows the largest deviation from the experimental data, and which is excluded from the fit. Because of the consistent pattern
shown by the $\chi^2(S)$ behavior, we assume that different models corresponding to intermediate values of the symmetry energy would lead to the $\chi^2$ value produced by the parabolic fit. Though this looks like a reasonable approximation, it will be important in the future to confirm these results with a full Bayesian estimation of the parameter space. To this aim, the development of an emulator-assisted inference framework is under construction. For further details of the protocol employed for the extraction of the symmetry energy constraint, refer to Ref.~\cite{Ciampi2025l}.
\subsection{Isospin diffusion current and probed baryonic density} \label{sec:probe}
As described in Section 1, isospin diffusion takes place through the neck region formed during the overlap of the projectile and target nuclei, where the baryonic density develops spatial gradients.
However, not all densities are explored with the same sensitivity; instead, different density regions contribute differently to the isospin diffusion process. Therefore, when attempting to constrain the density dependence of the symmetry energy, it is important to consider how strongly each density region influences the final observable quantities.\\
\indent
To better understand this, additional information about the reaction dynamics can be obtained from the BUU@VECC-McGill transport model simulation which can track how the reaction evolves over time and how different density regions are sampled quantitatively during the time when isospin diffusion currents are significant. The local particle density and velocity are evaluated by treating the test particles originating from the projectile ($P$) and target ($T$) nuclei separately prior to the collision. As described in Section \ref{Model}, the time evolution of the system is studied in the projectile frame, in which the target nucleus is boosted along the negative $z$ direction. The velocities of the $i^{\mathrm{th}}$ test particle in this frame are $\tilde{v}_{x,i}$, $\tilde{v}_{y,i}$ and $\tilde{v}_{z,i}$. In the centre-of-mass frame, the $x$  and $y$ components remain unchanged, while the $z$-component is shifted by the centre-of-mass velocity, i.e., $v_{z,i}^{cm}=\tilde{v}_{z,i}-v^{cm}$, where $v^{cm}$ depends on the masses of the projectile and target and the projectile beam velocity. Similarly, the positions of the particles are shifted by subtracting the centre-of-mass coordinates. The centre-of-mass position is obtained by averaging over all test particles. To evaluate the nucleon current, the particle positions and velocities are further transformed from the centre-of-mass frame to the principal-axis frame. In the principal-axis frame, the position and velocity vectors are expressed as $\vec{r_i} = x_i\hat{i} + y_i\hat{j} + z_i\hat{k}$ and $\vec{v_i} = v_{x,i}\hat{i} + v_{y,i}\hat{j} + v_{z,i}\hat{k}$, respectively. The transformation from the center-of-mass coordinates to the principal-axis coordinates is performed using a standard rotation:
\begin{eqnarray}
x_i=-z_i^{cm}\sin\theta+x_i^{cm}\cos\theta \; ; \; y_i=y_i^{cm} \; ; \; z_i=z_i^{cm}\cos\theta+x_i^{cm}\sin\theta\ \nonumber\\
v_{x,i}=-v_{z,i}^{cm}\sin\theta+v_{x,i}^{cm}\cos\theta; \; v_{y,i}=v_{y,i}^{cm} \; ; \; v_{z,i}=v_{z,i}^{cm}\cos\theta+v_{x,i}^{cm}\sin\theta
\end{eqnarray}
\begin{figure}[!b]
\begin{center}
\includegraphics[width=0.8\textwidth,keepaspectratio=true]{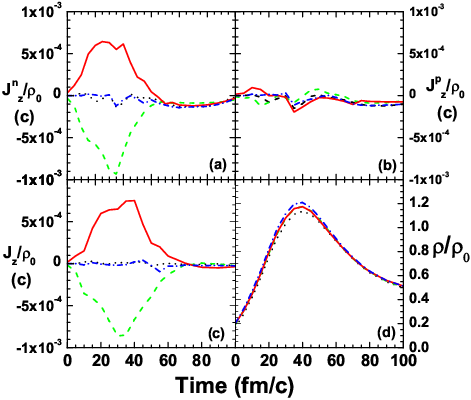}
\caption{Time dependence of neutron (upper left panel) and proton (upper right panel) current densities for the $^{58}$Ni+$^{58}$Ni (black dotted lines), $^{58}$Ni+$^{64}$Ni (green dashed lines), $^{64}$Ni+$^{58}$Ni (red solid lines) and $^{64}$Ni+$^{64}$Ni (blue dash-dotted lines) reactions at a projectile energy of $E_{p}$=32A MeV and an impact parameter $b$=5~$fm$. Lower left and right panels represent the net isospin current density and the probed baryonic density respectively. Theoretical calculations are performed using the \textit{ab initio-7} equation of state.}
\label{Isospin_Current1}
\end{center}
\end{figure}
\begin{figure}[!b]
\begin{center}
\includegraphics[width=0.8\columnwidth,keepaspectratio=true]{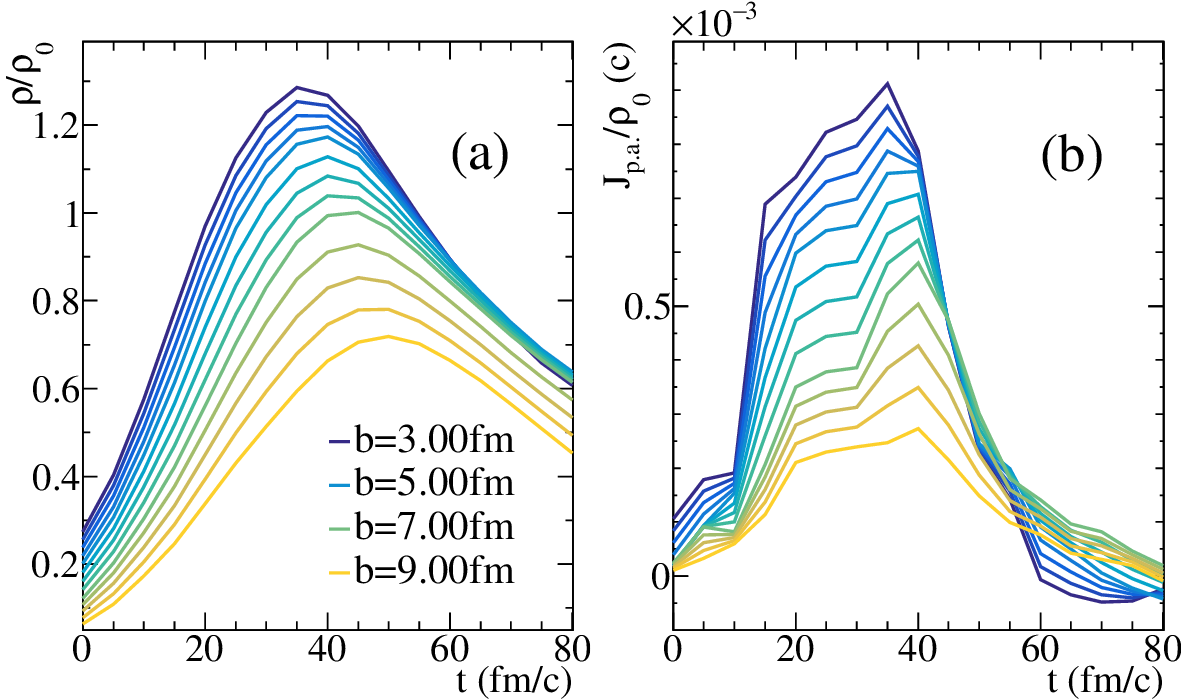}
\caption{Time evolution of (a) the normalized baryon density and (b) the isospin current density along the principal axis for the $^{64}$Ni$+^{58}$Ni at 32~MeV/nucleon, for impact parameters ranging from $b$= 3~fm to 9~fm. BUU@VECC-McGill calculations are performed with \textit{ab initio}~7 EoS,  for the impact parameters listed in the legend. Additional curves obtained via linear interpolation are plotted for impact parameter intervals of $\Delta b=0.5$~fm for a better representation}.
\label{Isospin_Current2}
\end{center}
\end{figure}
where $\theta$ represents the rotation angle between the center-of-mass frame and the principal-axis frame. This axis is determined at each time step by diagonalizing the event-averaged momentum of inertia tensor, defined as,
\begin{eqnarray}
I_{xx}=\sum_{i=1}^{N_{tot}}m\left \{(y_i^{cm})^2+(z_i^{cm})^2\right \} \; ; \; I_{zz}=\sum_{i=1}^{N_{tot}}m\left \{(y_i^{cm})^2+(x_i^{cm})^2\right \} \; ; \; I_{xz}=\sum_{i=1}^{N_{tot}}m x_i^{cm}z_i^{cm} \; ,
\end{eqnarray}
where $m$ denotes the nucleon mass, $N_{tot}=N_{test}(A_p+A_t)$ is the total number of test particles, $x_i^{cm}$, $y_i^{cm}$, and $z_i^{cm}$ represent the coordinates of the test particles in the centre of mass frame, and $(x,z)$ indicates the reaction plane.\\
\indent
The isospin diffusion current and the probed baryonic density within a small spherical volume in the neck region are evaluated at each time step. The radius of the sphere ($R$) is chosen such that it is not too large, to avoid contributions from spectator regions, and not too small, to limit statistical fluctuations. A representative value of $R = 3$ fm is used and stability of the results with respect to the choice of the $R$ parameter is checked. Further details regarding the choice of the radius are provided in Appendix A of Ref. \cite{Ciampi2025l}.\\
The configuration space is partitioned into a grid of cubic cells, with adjacent lattice points separated by a distance of  $l$ fm. Each cell therefore has a volume of $l^3$fm$^3$.
The proton and neutron current densities ($J^p_z$ and $J^n_z$ respectively) along the $z$ direction in the principal-axis frame are given by,
\begin{eqnarray}
J^p_z=\frac{1}{N_{grid}} \sum_{i=1}^{N_{grid}} \rho_{i,p} v_{z,i} \;\;\;\;\;\;\;\;\;\;\;\;\;\; J^n_z=\frac{1}{N_{grid}} \sum_{i=1}^{N_{grid}} \rho_{i,n} v_{z,i}
\end{eqnarray}

where $N_{grid}$ denotes the number of grid points inside the sphere. The proton density $\rho_{L,p}(\vec{r}_{\alpha})$ and neutron density $\rho_{L,n}(\vec{r}_{\alpha})$ at the lattice point $\vec{r}_{\alpha}$ are defined as
\begin{eqnarray}
\rho_{L,p}(\vec{r}_\alpha)=\sum_{i=1}^{N_{test}(Z_p+Z_t)}S(\vec{r}_{\alpha}-\vec{r}_i)
\;\;\;\;\;\;\;\;\;\;\;\;\;\;
\rho_{L,n}(\vec{r}_\alpha)=\sum_{i=1}^{N_{test}(N_p+N_t)}S(\vec{r}_{\alpha}-\vec{r}_i)
\end{eqnarray}

Similarly, the proton and neutron velocities ($v_{L,p}(\vec{r}_\alpha)$ and $v_{L,n}(\vec{r}_\alpha)$ respectively) at lattice point  $\vec{r}_{\alpha}$ are defined as
\begin{eqnarray}
\vec{v}_{L,p}(\vec{r}_\alpha)=\sum_{i=1}^{N_{test}(Z_p+Z_t)}\vec{v_i}S(\vec{r}_{\alpha}-\vec{r}_i)
\;\;\;\;\;\;\;\;\;\;\;\;\;\;
\vec{v}_{L,n}(\vec{r}_\alpha)=\sum_{i=1}^{N_{test}(N_p+N_t)}\vec{v_i}S(\vec{r}_{\alpha}-\vec{r}_i)
\end{eqnarray}

The index $\alpha$ represents the three coordinates of the lattice point, i.e., $\alpha = (x_l, y_m, z_n)$. The form factor is given by
\begin{equation}
S(\vec{r}) = \frac{1}{N_{test}(nl)^6} g(x) g(y) g(z)
\end{equation}
where
\begin{equation}
g(q) = (nl - |q|)\Theta(nl - |q|).
\end{equation}
In the present calculation, $l = 1$ fm and $n = 1$ are used. The isospin current density $J_{I,z}$  along the  are defined as\\
\begin{eqnarray}
J_{z}&=&J^n_{z}-J^p_{z}
\end{eqnarray}
\indent
Before discussing the current densities, it is useful to recall that isospin diffusion is initiated during the initial stage of the overlap between the projectile and target nuclei, when a low-density neck region is formed and nucleon exchange is facilitated. During the overlap stage, the density in the neck region increases and significant nucleon exchange takes place, leading to the development of a strong isospin current. As the reaction progresses, the projectile-like and target-like fragments separate, the neck breaks up, and both the nucleon exchange and the isospin current gradually vanish.\\
\indent
This evolution is illustrated in Fig. \ref{Isospin_Current1}, which shows the time dependence of the neutron and proton current densities through the neck region for the $^{58,64}$Ni+$^{58,64}$Ni reactions at a projectile energy of $E_{p}$=32A MeV and an impact parameter $b$=5 fm. As expected, in the case of an isospin symmetric system we do not observe any net current of protons or neutrons. Conversely, in the presence of a strong isospin gradient,
a sizeable neutron current is established during the contact time leading to a partial isospin equilibration, while the proton current is negligible.\\
\indent
The lower left panel shows that the isospin current $J_z$ reaches its largest magnitude during the interval $t\approx10$--70 fm/$c$, indicating that isospin diffusion is most active during this period. The lower right panel displays the corresponding evolution of the baryonic density within the neck region. Consequently, the densities explored during this time interval define the density domain most effectively probed by the isospin diffusion observable.\\
\indent
In order to define a weight function $w\big{(}\frac{\rho}{\rho_0}\big{)}$ that quantifies the relative influence of each explored density on the observed final isospin equilibration, it is necessary to consider how each density region contributes to the overall isospin diffusion process. The contribution of a given density region to the final observable depends not only on how long the reaction dynamics remain in that density region, but also on the magnitude of the isospin current that develops during that time interval. Therefore, both the time spent at a given density and the corresponding isospin transport strength must be taken into account when evaluating the importance of that density region. To incorporate these effects, the weight function at $b$=5~$fm$ is constructed by accumulating the baryonic density $\frac{\rho}{\rho_0}$ over time, with each time step weighted by the corresponding magnitude of the isospin current $J_{i}$, as given by:
 \begin{equation}
     w_{b=5}\Big{(}\frac{\rho}{\rho_0}\Big{)}=
     \int_{t_\text{start}}^{t_\text{stop}} J_{i}(t) \, \delta(\rho/\rho_0-\rho(t)/\rho_0)\,dt
 \end{equation}
The integration over time is performed from $t_\text{start}=0$~fm/$c$, corresponding to an initial distance of 12~$fm$ between the centres of the projectile and the target nuclei, up to $t_\text{stop}=80$~fm/$c$, at which time the contribution from the isospin diffusion current $J_{i}(t)$ becomes 
negligible. To calculate the total weight function over the range of impact parameters probed in the experiment, the time evolution of the relevant quantities was also extracted for three additional impact parameter values $b=3$, 5, 7 and 9~fm, with each case averaged over 200 events. The behaviour for intermediate impact parameters was obtained through interpolation, and the resulting distributions are presented in Fig. \ref{Isospin_Current2}.\\
\indent
To account for the contribution from different impact parameters, the partial weight functions, normalized by their integral over density and denoted as ${w}_b\big{(}\frac{\rho}{\rho_0}\big{)}$, are integrated over the selected impact parameter interval. The total weight function is therefore defined as
\begin{equation}
w\Big{(}\frac{\rho}{\rho_0}\Big{)} = \sum_{b=b_\text{min}}^{b_\text{max}}{w}_b\Big{(}\frac{\rho}{\rho_0}\Big{)}\, db
 \end{equation}
In this work, $b_\text{min}$=3~$fm$ and $b_\text{max}$=9~$fm$ are used. The likelihood function $\mathcal{L}\big(S(\frac{\rho}{\rho_0})\big)$, obtained from $\chi^2_{EoS}$ values calculated using the experimental ITR data points within this impact parameter interval, is then weighted by the function $w\big{(}\frac{\rho}{\rho_0}\big{)}$ evaluated over the same range.\\
\indent
Since the $\chi^2_{EoS}$ values are found to be minimum for the ab-initio-7 and SAMi equations of state, the same analysis was performed not only for the ab-initio-7 EoS but also for the SAMi EoS. The results obtained from the SAMi EoS calculations show a similar behaviour, consistent with the observations from the ab-initio-7 EoS calculations.
\begin{figure}[!b]
\centering
\includegraphics[width=0.65\textwidth,keepaspectratio=true]{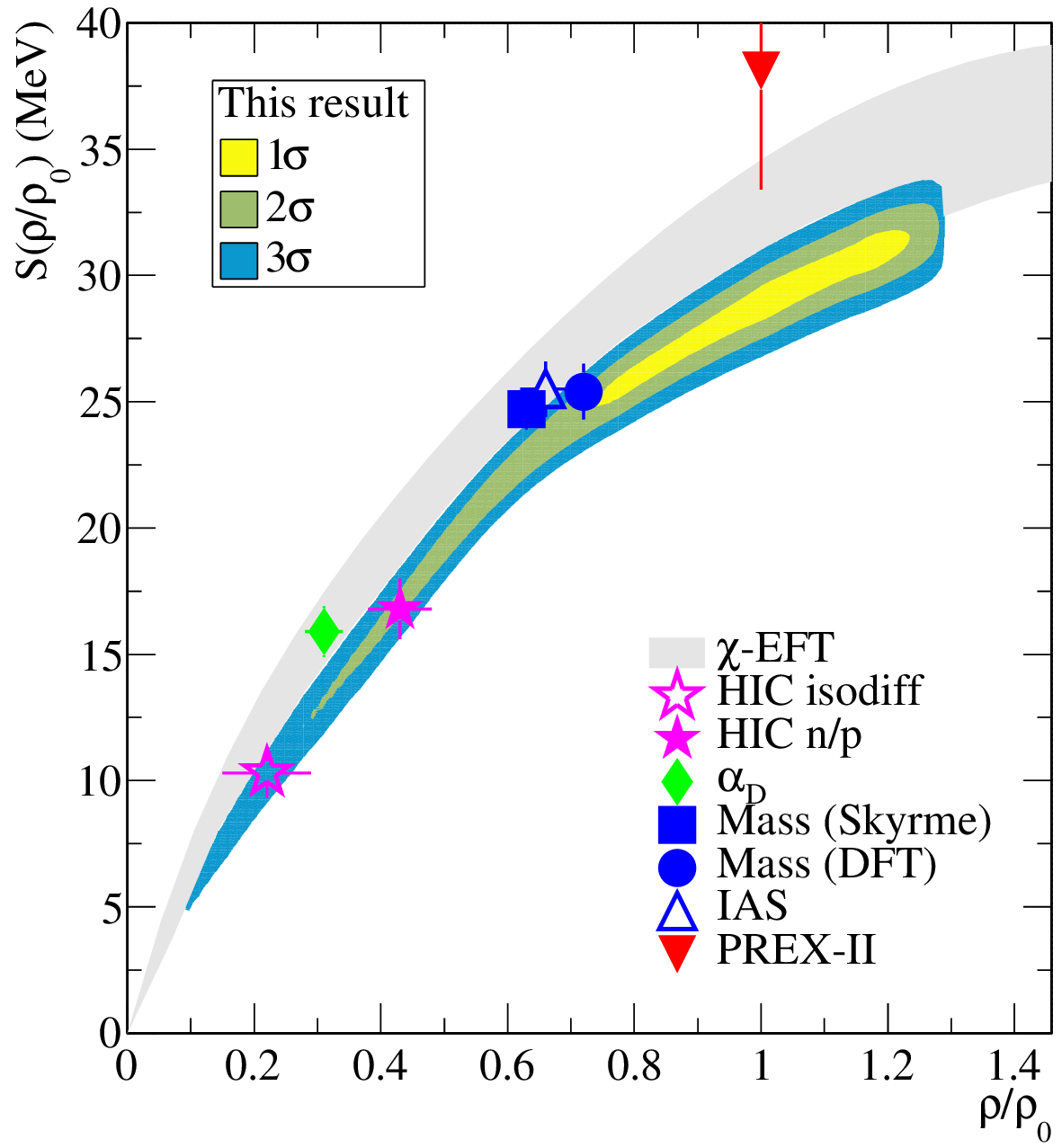}
\caption{Constraint on the density dependence of the symmetry energy extracted in the present work. The coloured regions represent the $1\sigma$, $2\sigma$, and $3\sigma$ confidence intervals, while the gray band corresponds to the microscopic uncertainty from \cite{Drischler}. Constraints from other observations available in the literature \cite{Lynch_2022,Tsang_2009,Morfouace_2019,Brown_2013,Kortelainen_2012,Danielewicz_2014,Zhang_2015,Reed} are also displayed for comparison.}
\label{Symmetry_energy_constraint}
\end{figure}
\subsection{Constraints on the Density Dependence of the Symmetry Energy}
The constraints on the density dependence of the symmetry energy are presented in Fig.~\ref{Symmetry_energy_constraint} in the form of confidence regions corresponding to the $1\sigma$ (yellow), $2\sigma$ (green), and $3\sigma$ (blue) levels.
For comparison, the 95$\%$ posterior distribution of $E_{sym}(\rho)$ that would be obtained without including the estimation of the density-sensitive region (\textit{i.e.}, by setting $w(\rho)=1$) is shown as a contour in the inset of Fig.~\ref{Symmetry_energy_constraint}. A good overall agreement is observed between the present results and the \textit{ab initio} calculations reported in Refs.~\citep{Drischler, Huth}. At the same time, tighter constraints are obtained within the density region effectively probed in the present study, and a preference towards the softer side of the uncertainty band derived from $\chi$-EFT and shown in gray is indicated.\\
\indent
It is emphasized that, in this work, both the extraction of the confidence regions for the symmetry energy and the determination of the baryonic densities probed through the analysis of isospin diffusion are performed in a consistent and unified manner within the same theoretical framework. It is also found that the sensitive density region lies close to the saturation density. As a consequence, stiff symmetry energy behaviors, such as those predicted by the NL3 model, can be excluded. Such stiff parametrizations are already known to be disfavored at high densities \citep{Sorensen_2024}, but they could not have been clearly discriminated if only observables probing densities below approximately $\rho_0/2$ had been considered.\\
\indent
A slight dependence of the extracted confidence regions on the chosen impact parameter interval and choice of the centrality radius $R$ is observed; however, it is found that the main conclusions remain stable against this arbitrary selection. Densities significantly below or above the interval $0.75 \lesssim \rho/\rho_0 \lesssim 1.15$, corresponding to the region constrained at the $1\sigma$ confidence level, are not effectively probed in the present experiment; consequently, no meaningful constraints can be extracted outside this density range. A comparison between the present results and the constraints obtained from various independent studies \cite{Lynch_2022}—namely, isospin diffusion data from Sn+Sn heavy-ion collisions \citep{Tsang_2009}, single and double neutron-to-proton spectral ratios \citep{Morfouace_2019}, analyses of nuclear binding energies \citep{Brown_2013,Kortelainen_2012}, isobaric analog states \citep{Danielewicz_2014}, electric dipole polarizability measurements $\alpha_D$ \citep{Zhang_2015,Rocamaza_2015}, and the neutron skin thickness \cite{Centelles_2009} of $^{208}$Pb measured in the PREX-II experiment \citep{Reed}—is also presented in Fig.~\ref{Symmetry_energy_constraint}. To obtain the density sensitive region of the different experimental probes shown in Figure ~\ref{Symmetry_energy_constraint}, the authors of ref.
\cite{Lynch_2022} proposed
two different methods, the slope and cross-over analysis, leading to a consistent estimation of the density region of highest sensitivity. These methods consist in determining the density at which the different functionals that reproduce the  observables within the experimental error bars present equivalent values of the symmetry energy, see ref.\cite{Lynch_2022} for details. We can see that these independent estimates are well consistent with our findings. In particular, a remarkable agreement is observed with previous results obtained from heavy-ion collisions, including isospin diffusion studies in Sn+Sn systems \citep{Tsang_2009}, with the main difference arising from the density sensitivity interval that has been identified.

\section{Conclusions}\label{Conclusions}
In this article, the isospin transport of the quasi-projectile formed in the $^{64,58}$Ni + $^{64,58}$Ni reactions at Fermi energies was investigated. The theoretical analysis was carried out within the BUU@VECC–McGill transport model framework. The isospin transport ratios were  extracted using two different observables. One observable is the neutron-to-proton ratio of the projectile remnant. The other observable is the neutron-to-proton ratio of forward-emitted free nucleons in the quasi-projectile reference frame. It was found that both observables are sensitive to the density dependence of the symmetry energy. However, different absolute values of the isospin transport ratios were obtained from these two observables, meaning that isospin ratios are not universal, and to construct them it is very important to choose an isospin sensitive variable that is accessible and reliable both in the experiment and in the transport model. It was also observed that the isospin transport ratio associated with the quasi-projectile shows stronger sensitivity to the nuclear equation of state than that obtained from free nucleons.\\
\indent
A model-independent experimental investigation of isospin equilibration was used as a benchmark for the present study. The impact parameter was reconstructed from charge particle multiplicities measured with the INDRA multidetector. The isotopic composition ($N/Z$) of the quasi-projectile was determined using the high-resolution FAZIA telescopes. The correlation between the reconstructed impact parameter and the quasi-projectile $N/Z$ provides a direct and model-independent measure of isospin diffusion. The experimentally extracted impact-parameter dependence of the isospin transport ratio was then compared with the predictions of the BUU transport model. The model calculations were performed using modern effective nuclear energy density functionals consistent with present constraints, including functionals extracted from ab-initio calculations with chiral interactions.
A detailed study of the time evolution of the baryon density during the reaction was performed. The time dependence of the isospin current density was also analyzed. From this combined analysis, the density region probed in the experiment was identified with improved reliability.\\
\indent
Stringent constraints on the density dependence of the symmetry energy term in the nuclear equation of state are obtained. The extracted posterior distribution of the symmetry energy parameters can be directly used in astrophysical applications.\\
\\

\textbf{Acknowledgements:}\\
This work has been partially supported by the IN2P3 Master Project MAC. S. Mallik acknowledges the GANIL Visiting Scientist program-2023 for very productive stay at Caen. S. Mallik also wishes to thank the VECC C$\&$I Group for providing high-performance computational facilities.\\
\textbf{Data Availability Statement:}\\
The supporting experimental data for this article are from the e789\_18 experiment and are registered as \cite{e789_data} following the GANIL Data Policy. The raw data supporting the conclusions of this article will be made
available by the authors on request.

\end{document}